\documentclass[a4paper]{article}
\usepackage{graphicx}
\usepackage{amssymb}
\usepackage{amsmath}

\begin{document}
\title{Pseudogap and the specific heat of high T$_c$ superconductors: a Hubbard model in a n-pole approximation}

\author{E. J. Calegari\footnote{eleonir@ufsm.br}, A. C. Lausmann\\
\it Laborat\'orio de Teoria da Mat\'eria Condensada\\
\it Departamento de F\'{\i}sica - UFSM, 97105-900\\
\it Santa Maria, RS, Brazil
\\
\\
S. G. Magalhaes\\
\it Instituto de F\'isica, Universidade Federal Fluminense\\
\it Av. Litor\^anea s/n, 24210, 346, Niter\'oi, Rio de Janeiro, Brazil
\\
\\
C. M. Chaves and A. Troper\\
\it Centro Brasileiro de Pesquisas F\'{\i}sicas\\
\it Rua Xavier Sigaud 150, 22290-180, Rio de Janeiro, RJ, Brazil}
%\ead{eleonir@ufsm.br}
\maketitle
\begin{abstract}
In this work the specific heat of a two-dimensional Hubbard model, suitable to discuss high-$T_c$ superconductors (HTSC),
 is studied taking into account hopping to first ($t$) and second ($t_2$) nearest neighbors.
% Results for the Hubbard model 
%show that the specific heat as a function of the temperature C(T) presents a two peaks structure [1]. The low temperature 
%peak has been associated with spin fluctuation while the high temperature peak is related to charge fluctuation. Nevertheless, 
%few attentions have been devoted to the interplay between the specific heat and the pseudogap . 
Experimental results for the 
specific heat of HTSC's, for instance, the YBCO and LSCO, indicate a close relation between the pseudogap and the specific heat.
 In the present work, we investigate the specific heat  by  the Green's function method within a $n$-pole approximation.
% proposed 
%by L. Roth [3,4]. 
The specific heat is calculated on the pseudogap and  on the superconducting regions.
%The analytical expression 
%for the specific heat has been obtained following the formalism presented in reference [5]. 
In the present scenario, the pseudogap emerges when the antiferromagnetic (AF) fluctuations
%correlations 
%(present in the Roth's band shift) 
become sufficiently strong. 
%to push down the 
%region of the nodal point (π,π) on the renormalized quasi-particle bands [7]. Superconductivity with   d x2 –y2 -   wave pairing is
% considered and the effects of AF correlations on superconductivity and on entropy, are investigated. 
The specific heat jump coefficient $\Delta \gamma$ decreases when the total occupation per site ($n_T$) reaches a given value. Such behavior of $\Delta \gamma$
indicates the presence of a  pseudogap in the regime of high occupation.
 
\end{abstract}

\section{Introduction}

The hole-doped high-$T_c$ cuprates are characterized by the presence of strong correlations in the underdoped regime where anomalies as the pseudogap are manifested \cite{Timusk}.
The normal state pseudogap observed in high-$T_c$'s has been intensively studied hoping to get information to help understand the unconventional 
superconductivity in such systems. The specific heat is a physical quantity that suffers the effects of the pseudogap. Indeed, experimental 
results for cuprates show that the specific heat is suppressed in the pseudogap region \cite{loram1,loram2}. This occurs because the pseudogap 
which develops near the Fermi energy depletes the spectral weight and affects the specific heat which is associated to the density of
states on the Fermi energy. The pseudogap occurs in the strong coupling regime \cite{Millis,Millis1,Millis2}, therefore, it is important to treat problem by using a theoretical technique that take into account the strong correlations
in an adequate way. However, much of the theoretical studies of the pseudogap regime are performed considering BCS-like  theories \cite{Tifrea,millan,Dzhumanov} which disregard correlations that 
might be important for the correct description of the pseudogap regime.

%The hopping to the second nearest neighbors 

In the present work, a two dimensional Hubbard model with hopping to first and second nearest neighbors is investigated by a $n$-pole approximation \cite{roth,edwards}. 
Such approximation preserves correlation functions for instance, the spin-spin correlation function $\langle \vec{S}_j\cdot\vec{S}_i\rangle$, that play an important role in the strong coupling regime. Furthermore, such $n$-pole approximation  allow us to consider
superconductivity with  $d_{x^2 -y^2}$-wave pairing and also to investigate the pseudogap regime of the two dimensional repulsive Hubbard model.

\section{Model and formalism}

The Hubbard model considered is described by the Hamiltonian:
\begin{equation}
\hat{H}=\sum_{\langle \langle ij \rangle\rangle \sigma} t_{ij}c_{i\sigma}^{\dag}c_{j\sigma} + \frac{U}{2}\sum_{i \sigma} n_{i,\sigma} n_{i,-\sigma}-\mu\sum_{i\sigma}n_{i\sigma}
\label{eqH1}
\end{equation}
where $c_{i\sigma }^{\dag }(c_{i\sigma })$ is the fermionic creation
(annihilation) operator at site $i$ with spin $\sigma
=\{\uparrow ,\downarrow \}$ and $n_{i,\sigma }=c_{i\sigma }^{\dag
}c_{i\sigma }$ is the number operator. The quantity $t_{ij}$ represents the hopping between sites $i$ and $j$
and $\langle \langle ...\rangle \rangle $ indicates the sum over the first
and second-nearest-neighbors of $i$. $U$ is the repulsive Coulomb potential between the $c$ electrons localized at the same site $i$ and $\mu $ the chemical
potential. The bare dispersion relation is  
%
%\begin{equation}
${\varepsilon }_{\vec{k}}=2t[\cos (k_{x}a) +\cos (
k_{y}a)] +4t_{2}\cos ( k_{x}a) \cos (k_{y}a)$ 
%\end{equation}
%
where $t$ is the first-neighbor and $t_{2}$ is the second-neighbor hopping
amplitudes.
%, $a$ is the lattice parameter and $t=1$ defines our unit of energy. 

The Green's function equation of motion  
\begin{equation}
\omega\langle\langle \hat{A};\hat{B}\rangle\rangle=\langle [\hat{A};\hat{B}]_+ \rangle+\langle\langle [\hat{A},\hat{H}];\hat{B}\rangle\rangle
\label{eqM}
\end{equation}
which relates the Hamiltonian $\hat{H}$ and the one-particle Green's function $\langle\langle \hat{A};\hat{B}\rangle\rangle$ presents the new Green's function 
$\langle\langle [\hat{A},\hat{H}];\hat{B}\rangle\rangle$. The equation of motion of this new Green's function is 
$\omega\langle\langle [\hat{A},\hat{H}];\hat{B}\rangle\rangle=\langle [[\hat{A},\hat{H}];\hat{B}]_+ \rangle+\langle\langle [[\hat{A},\hat{H}],\hat{H}];\hat{B}\rangle\rangle$ 
which involves a second new Green's function and implies that it is necessary to solve a infinite set of equations. An interesting way to treat this set of equations is 
%The Green's functions for the Hamiltonian introduced in equation (\ref{fig1}) is obtained within 
a $n$-pole approximation \cite{roth,edwards} which assumes that the commutator $[\hat{A},\hat{H}]$ can be rewritten as
\begin{equation}
[\hat{A}_n,\hat{H}]=\sum_m K_{nm}\hat{A}_m
\label{eqapr}
\end{equation}
in which $\{ \hat{A}_n\}$ is a set of operators describing the most important excitations of the system \cite{edwards}.
The quantity $\bf{K}$ present in equation \ref{eqapr} is determined from the relation $\bf{K}=\bf{E}\bf{N}^{-1}$
with $N_{nm}=\langle [\hat{A}_n;\hat{A}^{\dagger}_m]_+ \rangle$ and $E_{nm}=\langle [[\hat{A}_n;\hat{H}];\hat{A}^{\dagger}_m]_+ \rangle$.
In this way, the Green's function matrix is
\begin{equation}
\bf{G}(\omega)=\bf{N}(\omega\bf{N}-\bf{E})^{-1}\bf{N}.
\label{eqg}
\end{equation}

The Green's functions above contain the band shift $W_{\vec{k}\sigma}$ which can be written as
\begin{eqnarray}
W_{\vec{k}\sigma}&=&-\frac{1}{\langle n_{i\sigma}\rangle(1-\langle n_{i\sigma}\rangle)}\left\lbrace\sum_{\langle \langle ij \rangle\rangle} t_{ij}\langle c_{i\sigma}^{\dag}c_{j\sigma}(n_{i-\sigma}-n_{j-\sigma})\rangle\right.\\ \nonumber
& &\left.-\sum_{\langle \langle ij \rangle\rangle}e^{i\vec{k}\cdot(\vec{R}_j-\vec{R}_i)} t_{ij}\left [ \langle \vec{S}_j\cdot\vec{S}_i\rangle +\frac{1}{4}(\langle N_jN_i \rangle-\langle N_j\rangle\langle N_i\rangle) -\langle c^{\dagger}_{j\sigma}c^{\dagger}_{j-\sigma}c_{i-\sigma}c_{i\sigma} \rangle\right ]\right\rbrace
\label{eqWk}
\end{eqnarray}
where $N_j=n_{j\sigma}+n_{j-\sigma}$ is the total number operator. The correlation functions present in $W_{\vec{k}\sigma}$ can be determined self-consistently as in references \cite{roth,edwards,calegariEPJB}.
In particular, the spin-spin correlation function $\langle \vec{S}_j\cdot\vec{S}_i\rangle$ plays an important role in the present work because it is related to 
antiferromagnetic correlations which are one of the sources of a pseudogap \cite{Hiroaki,Tremblay}.

\subsection{Specific heat}

The specific heat is obtained from the energy per particle as $C=\frac{\partial E}{\partial T}$. 
%The energy per particle can be obtained from the Green's function 
Following the procedure described by Kishore and Joshi \cite{kishore},
 the energy per particle for the superconducting state of the Hubbard model introduced in equation (\ref{eqH1}) is:
\begin{equation}
 E_{S}=\frac{1}{2L}\sum_{\vec{k},\sigma}\sum_{i=1}^{4}Z_{i,\vec{k}\sigma}
 (\varepsilon_{\vec{k}}+\mu+E_{i,\vec{k}\sigma})f(E_{i,\vec{k}\sigma}) -\mu n_T\, 
\label{eqE} 
\end{equation}
where $n_T=n_{-\sigma}+n_{\sigma}$ is the total occupation, $Z_{i,\vec{k}\sigma}$ are the spectral weights \cite{edwards} of the Green's function 
$G^{(11)}_{\vec{k},\sigma}=\langle\langle c_{\vec{k},\sigma};c_{\vec{k},\sigma}^\dagger \rangle\rangle$ and $f(\omega)$ is the Fermi function.
In the superconducting state, 
%The $E_{i,\vec{k}\sigma}$ are the superconducting state 
the renormalized bands are:
%defined as:
%
\begin{equation}
 E_{i,\vec{k}\sigma}=(-1)^{(i+1)}\sqrt{\omega_{j,\vec{k}\sigma}^{2}+
 \, \frac{(-1)^{(j+1)}|\Delta_{\vec{k}}|^{2}[(\varepsilon_{\vec k}+Un_{-\sigma}-\mu)^{2}-\omega_{j,\vec{k}\sigma}^{2}]}{n_{-\sigma}^{2}(1-n_{-\sigma})^{2}(\omega_{2,\vec{k}\sigma}^{2}-\omega_{1,\vec{k}\sigma}^{2})}}\, ,
\end{equation}
with $j=1$ if $i=1$ or $2$, and $j=2$ if $i=3$ or $4$, $\Delta_{\vec{k}}=2t\Delta_0(cos(k_xa)-cos(k_ya))$ is the gap function where
$\Delta_0$ is the superconducting order parameter with $d_{x^2-y^2}$-wave symmetry\cite{edwards}.
In the normal state, the renormalized bands are:
\begin{equation}
 \omega_{j,\vec k\sigma}=\frac{U+\varepsilon_{\vec k}+W_{\vec k,\sigma}-2\mu}{2}-(-1)^{(j+1)}\frac{X_{\vec k,\sigma}}{2}
\end{equation}
where $X_{\vec{k}}=\sqrt{(U-\varepsilon_{\vec{k}}+W_{\vec{k}\sigma})^{2} +4\langle n_{-\sigma}\rangle U(\varepsilon_{\vec{k}}-W_{\vec{k}\sigma})}$.
%and $\varepsilon_{\vec{k}}$ is the unperturbed band energy
%\begin{equation}
%${\varepsilon }_{\vec{k}}=2t[\cos (k_{x}a) +\cos (
%k_{y}a)] +4t_{2}\cos ( k_{x}a) \cos (k_{y}a)$ 
%\end{equation}
%
%where $t$ is the first-neighbor and $t_{2}$ is the second-neighbor hopping
%amplitudes.
% and $a$ is the lattice parameter. 
%The $W_{\vec k,\sigma}$, is a band shift that depends on the 
%correlation function\cite{edwards,calegari} $\langle \vec{S}_i\cdot\vec{S}_j\rangle$.
\begin{figure}
\begin{center}
\includegraphics[angle=-90,width=10cm]{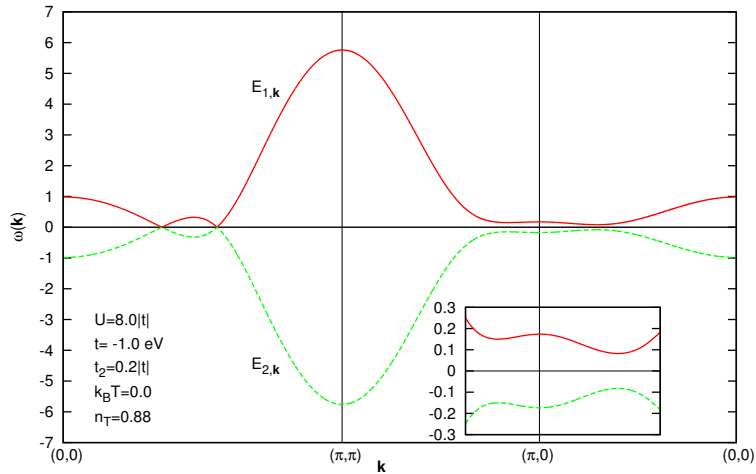}
\end{center}
\caption{The renormalized bands near the Fermi energy in the superconducting state. The inset shows in detail the gap
in the region of the $(\pi,0)$ point.}
\label{fig1}
\end{figure}

The specific heat jump is
\begin{equation}
\Delta C=[\frac{(C_S-C_N)}{C_N}]_{_{T=T_c}}
\label{DC}
\end{equation}
with $C_{S,N}=\frac{\partial E_{S,N}}{\partial T}$ in which
$E_S$ and $E_N$ are the energy per particle in the superconducting and in the normal sate, respectively.
%(given by equation (\ref{eqE})).
%and the normal state energy 
$E_N$ is
%also 
obtained from equation (\ref{eqE}), but keeping the superconducting order 
parameter being equal to zero $(\Delta_0=0)$.

\section{Results}

Figure \ref{fig1} show the bands close to the Fermi energy. Notice that the superconducting gap is present around the point $(\pi,0)$
but it is absent on the $k_x=k_y$ diagonal. This feature reflects the $d_{x^2-y^2}$-wave symmetry of the order parameter. 
The inset shows in detail the neighborhood of the $(\pi,0)$ point.

\begin{figure}
\begin{center}
\includegraphics[angle=-90,width=10cm]{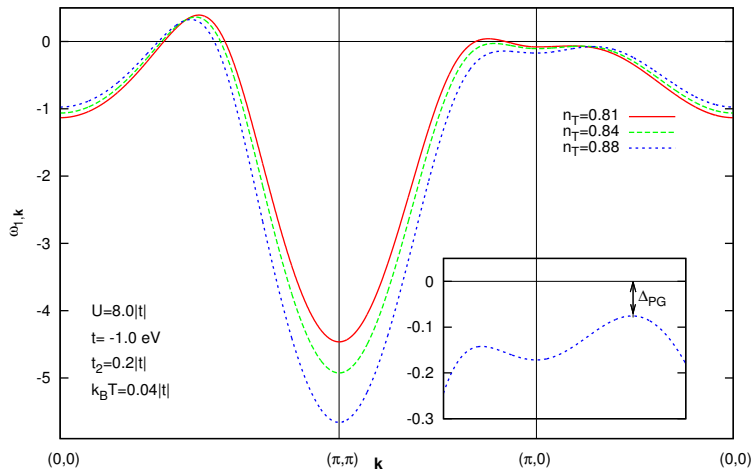}
\end{center}
\caption{The normal state renormalized  band $\omega_{1,\vec k\sigma}$ for different intensities of the total occupation $n_T$. The inset shows in detail the pseudogap  (for $n_T=0.88$) 
that appears in the  region around the $(\pi,0)$ point.}
\label{fig2}
\end{figure}

The normal state renormalized band $\omega_{1,\vec k\sigma}$ is shown in figure \ref{fig2} for different intensities of the total occupation $n_T$. 
%If nT increases a  pseudogap           develops in the          region. The inset shows explicitly the pseudogap        
%The renormalized quasiparticle band for the normal state is shown in figure \ref{fig2}. 
It is important to note that although the temperature $k_BT=0.04|t|$ is greater than $T_c(\lesssim 0.038|t| )$
a gap is still appearing around the point $(\pi,0)$. However, the gap appears only for $n_T$ above a given value, in this case $n_T\gtrsim 0.81$. A more carefully analysis show that 
the band crosses to above the chemical potential in the region of the $(\frac{\pi}{2},\frac{\pi}{2})$ point. This feature show that the gap developed in $(\pi,0)$
is actually a pseudogap like those observed in cuprate systems \cite{Timusk}.
The inset exhibit in details the pseudogap region (for $n_T=0.88$) and shows that the band does not reaches the chemical potential in the range of $\vec{k}$ where the pseudogap occurs.

%The bands are shifted down in the $(\frac{\pi}{2},\frac{\pi}{2})$  region. This feature is an effect of the antiferromagnetic correlations associated  with the spin-spin correlations               
%present in the band shift (see equation (\ref{eqWk})). 

\begin{figure}[h]
\includegraphics[width=20pc]{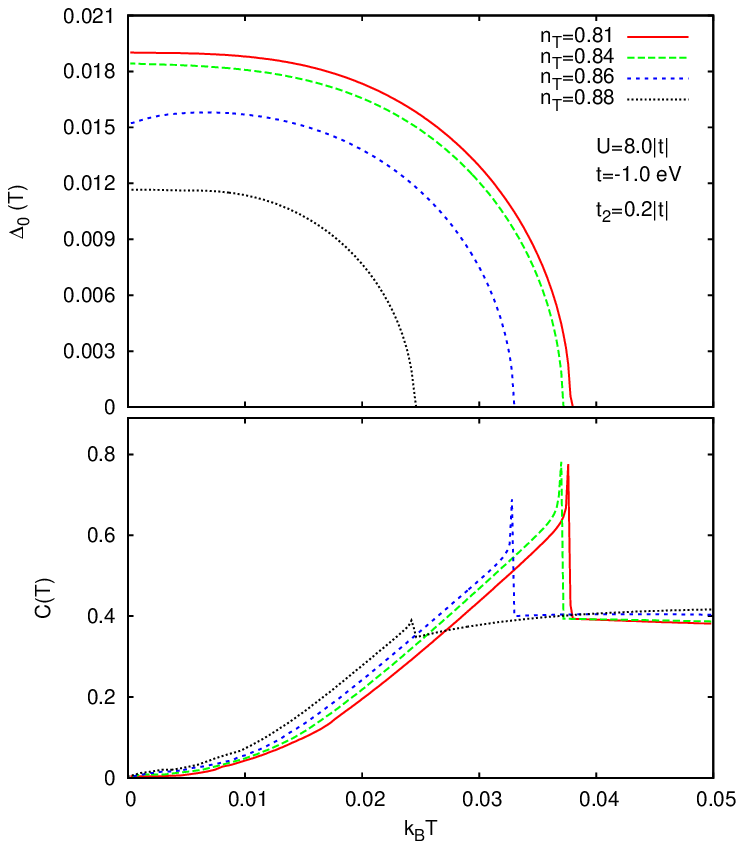}\hspace{2pc}%
\caption{\label{fig3} The upper panel shows the order parameter $\Delta_0(T)$ as a function of the temperature for several occupations $n_T$. The lower panel shows
the specific heat for the same parameters as in the upper panel. }
\end{figure}

\begin{figure}[h]
\begin{center}
\includegraphics[width=12cm]{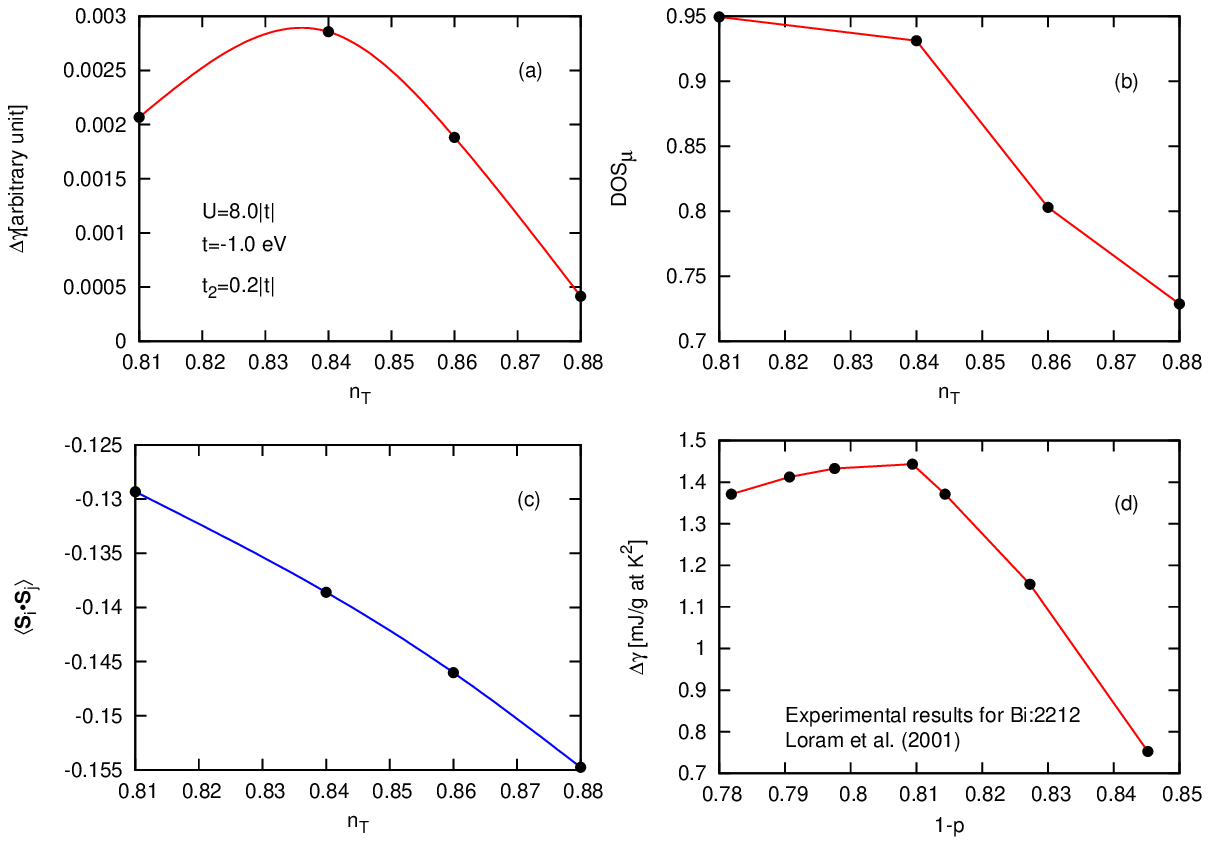}\hspace{2pc}%
\end{center}
\caption{\label{fig4} In (a) the specific heat jump coefficient $\Delta \gamma=\Delta C(T_c)/T_c$ as a function of the occupation $n_T$. The panels (b) and (c) show respectively the density of states on the chemical potential and
the spin-spin correlation function for the same model parameters as in (a). The panel (d) presents experimental results for the jump in the specific heat coefficient
$\Delta \gamma$
%=\Delta C(T_c)/T_c$ 
extracted from figure 3 of reference \cite{loram2}. p is the hole doping.}
\end{figure}

The upper panel in figure \ref{fig3} shows the order parameter $\Delta_0$ versus  the temperature for different occupations $n_T$.  The intensity of $\Delta_0$
decreases when $n_T$ increases leading the system to a regime of strong correlations \cite{Millis}. The lower panel displays the specific heat $C(T)$ for the same 
model parameters as in the upper panel. The anomaly on $C(T=T_c)$ is clearly observed. However, the most important feature in this result, is that, in the range of $n_T$ considered in this work, 
the specific heat jump on $T_c$ decreases when $n_T$ increases. In order to better understand such behavior, the specific heat jump coefficient $\Delta \gamma=\Delta C(T_c)/T_c$ 
( $\Delta C$ is defined in equation (\ref{DC})) 
as a function of the total occupation $n_T$ is shown in figure \ref{fig4}(a).
Notice that initially the $\Delta \gamma$ increases slightly with $n_T$ but, above $n_T\approx 0.835$, $\Delta \gamma$ starts to decrease. Such behavior for $\Delta \gamma$ is close related to the development 
of a pseudogap on the density of  states (DOS). The pseudogap suppresses the DOS at the chemical potential $\mu$ resulting in a reduction of the $\Delta \gamma$ which should be proportional to the DOS at $\mu$.
In figure \ref{fig4}(b) we present the  DOS at $\mu$ (DOS$_\mu$) calculated using the same parameters as in \ref{fig4}(a). The decreasing of DOS$_\mu$ with $n_T$ is directly related
to the pseudogap near de chemical potential and agrees with a high-resolution photoemission study \cite{ino} of La$_{2-x}$Sr$_x$CuO$_4$ which suggests that a pseudogap is the main
responsible for the similar behavior between the specific heat coefficient and the DOS$_\mu$ observed in the underdoped regime.

%The pseudogap is also observed in the quasiparticle band show in figure \ref{fig2}.

%

Figure \ref{fig4}(c) presents the spin-spin correlation function $\langle \vec{S}_j\cdot\vec{S}_i\rangle$ for the same model parameters as in \ref{fig4}(a). 
The $\langle \vec{S}_j\cdot\vec{S}_i\rangle<0$ is associated to large nearest-neighbor antiferromagnetic correlations. The modulus $|\langle \vec{S}_j\cdot\vec{S}_i\rangle|$
increases with $n_T$ evidencing the presence of strong antiferromagnetic fluctuations in the system. If the antiferromagnetic fluctuations are sufficiently strong it favors the appearance of a
pseudogap around the $(\pi,0)$ point (as show in figure \ref{fig2}) and the jump in the specific heat coefficient $\Delta \gamma$ starts to decrease.

The result show for $\Delta \gamma$ presented in \ref{fig4}(a) is in qualitatively agreement with experimental results for cuprates \cite{loram1,loram2}. 
In panel \ref{fig4}(d) we show the jump in the specific heat coefficient $\Delta \gamma$
%=C(T_c)/T_c$ 
extracted from figure 3 of reference \cite{loram2}. The quantity p indicates the number of 
holes per CuO$_2$ atoms.

\section{Conclusions}

The pseudogap and the superconducting regimes of a two-dimensional Hubbard model that includes hopping to first and second nearest neighbors is investigated.
The model has been investigated through a $n$-pole approximation \cite{roth,edwards} suitable to study the strong correlated regime.
The results show that even above $T_c$ a gap persists on the region of the $(\pi,0)$ and $(0,\pi)$ points in the first Brillouin zone which is a feature of a pseudogap with
$d_{x^2-y^2}$-wave symmetry. However, the pseudogap develops only above a given occupation $n_T$ where the antiferromagnetic fluctuations associated to the spin-spin correlation function 
$\langle \vec{S}_j\cdot\vec{S}_i\rangle$ becomes sufficiently strong. 
Such pseudogap affects the specific heat suppressing the jump in the specific heat coefficient $\Delta \gamma$. 
The result for $\Delta \gamma$ is in qualitatively agreement with experimental results for some cuprates \cite{loram1,loram2}.
In summary, in this work we report some results obtained by treating the Hubbard model within an approximation \cite{roth,edwards} that is adequate for investigate the strong coupling regime
of the model. The obtained results are in qualitatively agreement with some experimental results \cite{loram1,loram2} and are consistent with the recent dynamical cluster approximation (DCA) 
studies \cite{Millis,Millis1,Millis2} 
%for the repulsive two-dimensional Hubbard model 
which asserts 
%In the present scenario it has been 
%observed 
that the pseudogap at issue is a feature of the strongly coupling regime.
%agreeing
%with the recent dynamical cluster approximation (DCA) studies \cite{Millis,Millis1,Millis2} for the repulsive two-dimensional Hubbard model.

%\subsection{Acknowledgments}
%Authors wishing to acknowledge assistance or encouragement from 
%colleagues, special work by technical staff or financial support from                                                                                                                                                                                                                                                                                       
%organizations should do so in an unnumbered Acknowledgments section 
%immediately following the last numbered section of the paper. The 
%command \verb"\ack" sets the acknowledgments heading as an unnumbered
%section.

\section{Acknowledgment}
This work was partially supported by the Brazilian agencies CNPq,
%(Conselho Nacional de Desenvolvimento Cient\'{\i}fico e Tecnol\'ogico),
CAPES and
%(Coordena\c{c}\~ao de 
%Aperfei\c{c}oamento de Pessoal de N\'{\i}vel Superior), 
FAPERGS.
%(Funda\c{c}\~ao de Amparo \`a Pesquisa do Rio Grande do Sul) 
%and FAPERJ.
%(Funda\c{c}\~ao Carlos Chagas Filho de Amparo \`a Pesquisa do Estado do Rio de Janeiro).

%\section* {References}

\end{document}